\documentclass{article}
\usepackage[T1]{fontenc} 
\usepackage[utf8]{inputenc} 
\usepackage{ismir,amsmath,cite,url}
\usepackage{graphicx}
\usepackage{float}

\usepackage{color}

\usepackage[firstpage]{draftwatermark}
\definecolor{lightgray}{rgb}{0.9,0.9,0.9}
\definecolor{darkgray}{rgb}{0.4,0.4,0.4}
\SetWatermarkFontSize{12pt}
\SetWatermarkScale{1.1}
\SetWatermarkAngle{90}
\SetWatermarkHorCenter{202mm}
\SetWatermarkVerCenter{170mm}
\SetWatermarkColor{darkgray}
\SetWatermarkText{Late-Breaking / Demo Session Extended Abstract, ISMIR 2024 Conference}

\def\lbd{}	

\title{Exploring Transformer-Based Music Overpainting for Jazz Piano Variations}

\multauthor
{Eleanor Row$^1$ \hspace{1.2cm} Ivan Shanin$^1$ \hspace{1.2cm} George Fazekas$^1$} 
{$^1$ Centre for Digital Music, Queen Mary University of London, UK \\
{\tt \{e.r.v.row, ivan.shanin\}@qmul.ac.uk}}

\def\authorname{E. Row, I. Shanin, and G. Fazekas}

\begin{document}

\maketitle

\begin{abstract}

This paper explores transformer-based models for music overpainting, focusing on jazz piano variations. Music overpainting generates new variations while preserving the melodic and harmonic structure of the input. Existing approaches are limited by small datasets, restricting scalability and diversity. We introduce VAR4000, a subset of a larger dataset for jazz piano performances, consisting of 4,352 training pairs. Using a semi-automatic pipeline, we evaluate two transformer configurations on VAR4000, comparing their performance with the smaller JAZZVAR dataset. Preliminary results show promising improvements in generalisation and performance with the larger dataset configuration, highlighting the potential of transformer models to scale effectively for music overpainting on larger and more diverse datasets.

\end{abstract}

\section{Introduction}\label{sec:introduction}

Music overpainting is a generative task that creates variations of a musical excerpt, with the aim of preserving its melodic and harmonic structure \cite{rowJAZZVARDatasetVariations2023, rowAdvancingAIMusic2024}. Models trained for this task have the potential to be adopted as GenAI (Generative AI) creative tools in music composition, enabling composers to generate new variations from existing musical excerpts. However, overpainting has primarily been explored using small-scale datasets, and there has been limited research into how well models can generalise to larger datasets, largely due to the scarcity of data for the task.

To address this gap, we are developing a large-scale structured dataset for solo piano jazz performances, aligned with lead sheets, as described in \cite{shaninAnnotatingJazzRecordings2023}. This dataset aligns lead sheets to both the head and solo sections of these jazz performances, providing more data for exploring Music Information Retrieval (MIR) tasks, such as structural analysis and music generation. Existing jazz datasets like JAAH and JSD provide structural annotations but focus on solo sections or multi-instrument performances \cite{ eremenkoJAAHAudioalignedJazz2018, balkeJSDDatasetStructure2022}. The Weimar Jazz Database offers solo excerpts but lacks specificity for solo piano performances \cite{Pfleiderer:2017:BOOK}. The dataset we are developing aims to overcome these limitations by providing detailed, structured data for solo piano jazz analysis. Additionally, we are developing an automatic data collection and structuring pipeline to facilitate the creation of large-scale datasets in general, which can be extended beyond jazz to other musical genres.

As part of our work on the large-scale dataset project, we developed a semi-automatic pipeline to extract ‘variations’ from jazz piano performances, by aligning small sections from lead sheets with corresponding performances from the head section, based on the method described in \cite{rowJAZZVARDatasetVariations2023}. This semi-automatic pipeline led to the creation of VAR4000, a subset of the larger dataset project. In this Late-Breaking Demo, we use VAR4000 to apply a transformer-based model to the music overpainting task, exploring how the model performs on a larger and more diverse dataset \cite{huang2018music}.

In this paper, we present preliminary work that investigates how different configurations of a transformer architecture influence model performance when applied to a larger dataset, compared to the JAZZVAR dataset. Our aim is to work towards establishing a new baseline for music overpainting, particularly for future studies involving larger and more diverse datasets, and to assess the model's ability to generalise to these datasets. We explore two model configurations on the VAR4000 dataset, comparing their performance and scalability. Our preliminary results show how model performance varies across configurations, providing insights into potential improvements for handling larger datasets and enhancing model robustness.

\section{Methodology}\label{sec:methodology}

\begin{table*}[t]
\begin{center}
\begin{tabular}{|l|c|c|c|c|}
\hline
\textbf{Feature} & \textbf{JAZZVAR Originals} & \textbf{JAZZVAR Variations} & \textbf{VAR4000 Originals} & \textbf{VAR4000 Variations} \\
\hline
Pitch Class Entropy & 2.94 ± 0.24 & 3.13 ± 0.24 & 3.05 ± 0.25 & 3.26 ± 3.26 \\
\hline
Pitch Range         & 36.44 ± 3.60 & 10.91 ± 10.91 & 33.00 ± 3.32 & 47.00 ± 10.70 \\
\hline
Polyphony           & 5.30 ± 0.28  & 2.08 ± 2.08   & 4.69 ± 0.33  & 4.88 ± 1.57 \\
\hline
No. of Pitches      & 16.08 ± 0.28 & 8.05 ± 8.05   & 18.00 ± 4.15 & 32.00 ± 8.53 \\
\hline
Pitch in Scale      & 0.89 ± 0.24  & 0.08 ± 0.08   & 0.88 ± 0.09  & 0.80 ± 0.07 \\
\hline
\end{tabular}
\end{center}
\caption{Comparison of various features between JAZZVAR (505 data pairs) and VAR4000 (4,352 data pairs) datasets, with mean and standard deviation for both Originals and Variations.}
\label{tab:comparison}
\end{table*}

\subsection{Data}

We use a subset of a large-scale dataset we are compiling for music generation and music information retrieval tasks. This subset consists of 4,352 pairs of ‘Original’ and ‘Variation’ MIDI data. Similar to the JAZZVAR dataset, the ‘Original’ segments are 4-bar melody and chord excerpts taken from a lead sheet transcription of a jazz standard. The ‘Variation’ segments are extracts from audio jazz piano performances, which are semi-automatically aligned to the ‘Original’ segments using Viterbi decoding \cite{shaninAnnotatingJazzRecordings2023}. The deep chroma features of the audio are compared to the chord symbols from the lead sheet. Using these alignments, we extract the corresponding MIDI transcriptions from the PiJAMA dataset \cite{edwardsPiJAMAPianoJazz2023}. These segments are then verified through human evaluation. For ease of reference, we will refer to this subset as VAR4000. We perform the same data transposition and augmentation as in \cite{rowAdvancingAIMusic2024} on VAR4000. This process increased the sample size from 4,352 pairs to 52,224 pairs. The data was tokenised with RemiPlus \cite{rutteFIGAROControllableMusic2022}. In comparison, JAZZVAR contains 505 data pairs, and when augmented the sample size increased to 6,060 pairs.

\subsection{Training Setup}

We explored two configurations of a transformer-based model architecture \cite{huang2018music}. Model 1 consisted of 2 layers, a hidden dimension of 64, 8 attention heads, and a feed-forward dimension of 256. Model 2 had 4 layers, a hidden dimension of 128, 8 attention heads, and a feed-forward dimension of 512. Both models were optimised using Adam with an initial learning rate of 1e-3, adjusted via a learning rate scheduler, and a batch size of 16. Early stopping was applied, with Model 1 training for 131 epochs on JAZZVAR and 80 epochs on VAR4000, while Model 2 trained for x epochs. All models were trained on one NVIDIA RTX A5000, and nucleus sampling was used to generate the outputs \cite{bjareExploringSamplingTechniques2023}.

\section{Preliminary Experiments and Discussion}\label{sec:evaluation}

The musical feature evaluation metrics, including Pitch Class Entropy (PCE), Pitch Range (PR), Polyphony (P), Number of Pitches (NoP), and Pitch in Scale (PS), were calculated for both the JAZZVAR and VAR4000 datasets to understand their distribution \cite{dongMuseganMultitrackSequential2018a}. As shown in Table \ref{tab:comparison}, VAR4000 Variations exhibit higher PCE and a significantly larger Pitch Range than JAZZVAR Variations, suggesting greater pitch diversity and broader musical variation. Polyphony remains consistent between the datasets, with VAR4000 Variations slightly more polyphonic.

We trained the JAZZVAR dataset using Model 1 and applied the same configuration to VAR4000 for comparison. Table \ref{tab:abbreviatedmodels} shows the feature metric comparison across the entire test set. Notably, the outputs from Model 2 are closer to the manually annotated JAZZVAR variations, suggesting Model 2's results align more closely with human-generated variations.

Five outputs from each model’s test set were also qualitatively evaluated by listening to assess how well the generated outputs retained the melody and harmony of the primer input. 
Preliminary results indicate that VAR4000 outperformed JAZZVAR in Model 1, demonstrating better scalability to the larger dataset. Model 2, tested exclusively on VAR4000, performed better than both JAZZVAR and VAR4000 in Model 1, showing that a more complex architecture significantly improved performance.

These findings highlight the potential for enhanced scalability with more complex transformer architectures, particularly as access to larger datasets increases. Additionally, Model 2 exhibited better generalisation to VAR4000, which is critical for real-world applications of GenAI in music composition, where users are likely to input diverse musical material into the model. An improved evaluation pipeline was also implemented to mitigate data leakage, providing a more reliable measure of the model's ability to generalise to unseen data.



\begin{table}[ht]
\begin{center}
\begin{tabular}{|l|c|c|c|}
\hline
\textbf{Feat.} & \textbf{M1-JV} & \textbf{M1-V4} & \textbf{M2-V4} \\
\hline
PCE & 3.21 ± 0.26 & 3.34 ± 0.22 & 2.66 ± 0.66 \\
\hline
PR & 46.95 ± 9.16 & 53.07 ± 11.23 & 33.24 ± 12.54 \\
\hline
P & 5.11 ± 1.06 & 3.03 ± 2.17 & 3.71 ± 1.65 \\
\hline
NoP & 29.49 ± 9.43 & 38.41 ± 10.55 & 13.56 ± 7.37 \\
\hline
PS & 0.80 ± 0.07 & 0.77 ± 0.09 & 0.86 ± 0.10 \\
\hline
\end{tabular}
\end{center}
\caption{Comparison of musical features between Model 1 trained on the JAZZVAR dataset (M1-JV) and the VAR4000 dataset (M1-V4), and Model 2 trained on the VAR4000 dataset (M2-V4) with mean and standard deviation.}
\label{tab:abbreviatedmodels}
\end{table}
\section{Conclusion and Future Work}\label{sec:evaluation}

In conclusion, we investigated the performance of a transformer-based model on VAR4000, a subset of a larger dataset that we are currently working on, in comparison to the smaller JAZZVAR dataset. While the results are promising, further work is needed to improve the scalability and generalisation of the model. Expanding the dataset and experimenting with additional configurations will help optimise performance for larger datasets.

We plan to explore a custom loss function to enhance output quality by focusing more on the original segments during training. Improved evaluation metrics are also necessary to better assess model performance. Future work will include subjective evaluations with composers and jazz experts to refine the model and explore its potential beyond jazz.

\section{Acknowledgments}\label{sec:acknowledgements}

Eleanor Row is a research student at the UKRI Centre for Doctoral Training in Artificial Intelligence and Music, supported by UK Research and Innovation [grant number EP/S022694/1].

\bibliography{latex/ISMIR_LBD_BIB}

%
%
%
%
%

\end{document}